\documentclass
[prl,twocolumn,superscriptaddress,floatfix,showpacs,10pt]{revtex4}%
\usepackage{amsfonts}
\usepackage{amsmath}
\usepackage{amssymb}
\usepackage{graphicx}%
\begin{document}
\title{Zwei-Dreibein Gravity}

\author{Eric A. Bergshoeff}
\email{e.a.bergshoeff@rug.nl}
\affiliation{Centre for Theoretical Physics, University of Groningen, Nijenborgh 4, 9747 AG Groningen, The Netherlands}

\author{Sjoerd  de Haan}
\email{s.de.haan@rug.nl }
\affiliation{Centre for Theoretical Physics, University of Groningen, Nijenborgh 4, 9747 AG Groningen, The Netherlands}

\author{Olaf Hohm}
\email{olaf.hohm@physik.uni-muenchen.de}
\affiliation{Arnold Sommerfeld Center for Theoretical Physics, Theresienstrasse 37, D-1-80333 Munich, Germany}

\author{Wout Merbis}
\email{w.merbis@rug.nl}
\affiliation{Centre for Theoretical Physics, University of Groningen, Nijenborgh 4, 9747 AG Groningen, The Netherlands}

\author{Paul K. Townsend}
\email{p.k.townsend@damtp.cam.ac.uk}
\affiliation{Department of Applied Mathematics and Theoretical Physics, Centre for Mathematical Sciences, University of Cambridge,
Wilberforce Road, Cambridge, CB3 0WA, U.K.}

\begin{titlepage}
\begin{flushright}  DAMTP-2013-34, LMU-ASC 47/13, RUG-2013-26 %Preprint-number
%$\hspace{2.1cm}{}$
\end{flushright}
\vfill

\end{titlepage}

\begin{abstract}

We present a generally-covariant and parity-invariant ``zwei-dreibein''  action for gravity in three space-time dimensions that  propagates  two massive spin-2 modes, unitarily, and we use Hamiltonian methods to confirm the absence of unphysical degrees of freedom. We show how zwei-dreibein gravity  unifies previous   ``3D massive gravity''  models, and extends them, in the
context of the AdS/CFT correspondence, to  allow for a positive central charge consistent with bulk unitarity.

\end{abstract}

\maketitle

\setcounter{equation}{0}

Einstein's theory of General Relativity (GR) can be viewed as a field theory describing the interactions of a massless spin-$2$ particle, the graviton. This point of view suggests
that GR might be generalized to allow for a small graviton mass. The natural starting point for such investigations is the massive spin-$2$ field theory constructed long ago by Fierz and Pauli,  but it has proved difficult to find a consistent interacting version of this theory. One problem is the generic appearance of an unphysical scalar mode of negative energy: the Boulware-Deser ghost
\cite{Boulware:1973my}. In the context of a three-dimensional (3D) spin-$2$ theory, many of the problems were  resolved a few years ago by the construction of ``New Massive Gravity'' (NMG).  Although this is a general covariant (diffeomorphism invariant) theory, of
4th order in derivatives in its initial formulation, linearization yields a free field theory that is equivalent to the 3D version of the Fierz-Pauli (FP) massive spin-$2$ theory  \cite{Bergshoeff:2009hq}. Moreover, the non-linearities are precisely those for which the Boulware-Deser ghost is avoided \cite{deRham:2011ca}.

In an initially parallel development, a ghost-free non-linear extension of the four-dimensional (4D) FP spin-$2$ field theory was constructed by de Rham, Gabadadze and Tolley \cite{deRham:2010kj}.  One unattractive feature  of this ``dRGT'' model is the fact that it involves a  fixed background metric, in addition to the dynamical metric, and so lacks the general covariance of Einstein's General Relativity.
General covariance can be restored, albeit at the cost of re-introducing a massless graviton, by considering an alternative  ``bi-metric''  gravity model \cite{Hassan:2011zd}  from which the dRGT model can be recovered as a truncation in which one metric is taken to be non-dynamical.
The structure of  this  bi-metric model   becomes quite  simple when formulated in terms of two vierbeins rather than two metrics \cite{Hinterbichler:2012cn}; i.e.~when formulated as a ``zwei-vierbein'' model  (see \cite{Chamseddine:1978yu} for earlier related  models).
In this formulation the absence of ghosts can be seen easily, although it has been shown
that shock waves in these 4D massive gravity models propagate acausally \cite{Deser:2012qx,Deser:2013gpa,Deser:2013eua}.

Whereas the motivation for 4D massive gravity comes mainly from potential applications to cosmology, the motivation for 3D massive gravity stems from simplifying features of the quantum theory arising from the lower dimension. In particular, 3D massive gravity models provide a new arena for  the AdS/CFT correspondence, in which a 3D quantum gravity theory in an asymptotically anti-de Sitter (AdS) space-time is conjectured to be  equivalent to a 2D conformal field theory (CFT) on the AdS boundary.  In the semi-classical approximation to the quantum gravity theory  the dual  CFT has a large central charge that can be computed from the asymptotic symmetry algebra \cite{Brown:1986nw}.  The AdS/CFT correspondence does not obviously apply to the 3D dRGT model because of its lack of general covariance. It does apply to NMG but perturbative unitarity (in the bulk) holds if and only if  the central charge  of the boundary CFT is negative, which implies that the CFT is non-unitary.

This clash between bulk and boundary unitarity is a feature of all currently known  generally-covariant  3D models of massive gravity.
In particular, the bi-metric model of \cite{Banados:2009it}, although permitting a positive central charge, is not unitary in the bulk due to the Boulware-Deser ghost.  In this paper we present a parity-preserving ``zwei-dreibein'' model of massive gravity that overcomes this problem. The general model of zwei-dreibein gravity (ZDG) has five continuous parameters, and a choice of sign,  but we find a parameter range for which the bulk theory is  unitary and has a boundary CFT with positive central charge. For other choices of the parameters we exhibit limits in which the dRGT model and NMG are recovered  \cite{Hassan:2011zd,Paulos:2012xe}, the latter in its ``Chern-Simons-like'' form \cite{Hohm:2012vh}.  ZDG  thus unifies these two rather different approaches to massive gravity in three dimensions, in addition to extending them in a way that resolves a major difficulty.

The ZDG fields are a pair $\{e_I^a; I=1,2; a=0,1,2\}$ of Lorentz vector valued one-forms, and a pair $\omega_I^a$ of Lorentz-vector valued connection one-forms, from which we may construct pairs of torsion and curvature 2-forms:
\begin{equation}
T_I^a = d e_I^a + \varepsilon^{abc} \omega_{I\, b} e_{I\, c}\, , \quad R_I^a = d\omega_I^a + \frac{1}{2} \varepsilon^{abc} \omega_{I\, b} \omega_{I\, c}\, .
\end{equation}
We use a notation in which the exterior product of forms is implicit.
It will  be convenient to introduce a sign $\sigma =\pm1$ and two independent positive mass parameters $M_I$,  and to define \cite{Paulos:2012xe}
\begin{equation}
M_{12} = \left(\sigma M_1 M_2\right)/\left(\sigma M_1 + M_2\right)\, .
\end{equation}
This  is positive for $\sigma=1$,  and finite for finite $M_I$.  It may have either sign when $\sigma=-1$, and in this case we assume that $M_1\ne M_2$.  The Lagrangian 3-form for ZDG can now be written as
\begin{equation}
{\cal L}_{\rm ZDG} =  {\cal L}_{1} + {\cal L}_{2} + {\cal L}_{12}\, ,
\end{equation}
where ${\cal L}_1$ and ${\cal L}_2$ are Einstein-Cartan (EC) Lagrangian 3-forms; i.e.
\begin{equation}\label{LI}
\begin{split}
{\cal L}_{1} &= - \sigma M_1 e_{1\, a} R_1^a - \frac{1}{6} m^2 M_1 \alpha_1 \varepsilon_{abc} e_1^a e_1^b e_1^c \, , \\
{\cal L}_{2} &= - M_2 e_{2\, a} R_2^a - \frac{1}{6} m^2 M_2 \alpha_2 \varepsilon_{abc} e_2^a e_2^b e_2^c \, ,
\end{split}
\end{equation}
where $m$ is a further mass parameter and $\alpha_I$ are two dimensionless ``cosmological''  parameters.
The two EC terms are coupled by the third term
\begin{equation}
{\cal L}_{12} =  \frac{1}{2} m^2 M_{12}\, \varepsilon_{abc} \left( \beta_1 e_1^a e_1^b e_2^c + \beta_2 e_1^a e_2^b e_2^c\right) \, ,
\end{equation}
where $\beta_I$ are two dimensionless parameters.

The general ZDG model, as constructed above, depends on  five independent continuous parameters: $(\alpha_I,\beta_I)$ and the ratio $M_1/M_2$. The mass parameter $m$ is convenient but  inessential because we  could consider $m^2(\alpha_I,\beta_I)$ as independent mass-squared parameters from which four dimensionless parameters can be  found by taking ratios with, say, $M_1M_2$.  In the absence of the ${\cal L}_{12}$ term  the action  is the sum of  two EC actions and so has diffeomorphism and local Lorentz gauge invariance separately for the two  sets of EC form fields; this is broken by ${\cal L}_{12}$ to the  diagonal EC  gauge invariance found by identifying the two sets of EC gauge parameters.

We now expand the two dreibeine  about a common fixed dreibein $\bar e^a$ of a maximally-symmetric  background with cosmological constant $\Lambda$,  and similarly for their respective spin connections:
\begin{equation}
\begin{split}
e_1^a &=   \bar e^a + \kappa h_1^a \, , \qquad \;  \omega_1^a = \bar\omega^a + \kappa v_1^a \, , \\
e_2^a &=   \gamma(\bar e^a + \kappa h_2^a) \, , \quad  \omega_2^a = \bar\omega^a + \kappa v_2^a \, ,
\end{split}
\end{equation}
where $\gamma$ is a constant and $\kappa$ is a small expansion parameter.
The ZDG action may now be expanded in powers of $\kappa$. Cancellation of the linear terms both fixes $\Lambda$,
\begin{equation}\label{fixLam}
\Lambda/m^2 = - \gamma^2\alpha_2 + (M_{12}/M_2) \left(\beta_1 + 2\gamma \beta_2\right)\, ,
\end{equation}
and  imposes the following quadratic constraint  on $\gamma$\,:
\begin{eqnarray}\label{quadgam}
&&\left[\alpha_2 \left(\sigma M_1 + M_2\right) + \beta_2 M_2\right]\gamma^2 + 2 \left(M_2\beta_1-\sigma M_1\beta_2\right)\gamma \nonumber \\
&&\  -\,  \sigma\left[\alpha_1\left(\sigma M_1 + M_2\right) + \beta_1 M_1\right] =0\, .
\end{eqnarray}

The  quadratic terms in the expansion of the action may now be diagonalised,
provided
 \begin{equation}
  M_{\rm crit} = \sigma M_1+\gamma M_2 \neq 0\;,
 \end{equation}
by introducing the new one-form fields
\begin{eqnarray}
h_+^a &=& \left(\sigma M_1 h_1^a +  \gamma M_2 h_2^a\right)/M_{\rm crit}\, ,\nonumber \\
 v_+^a &=& \left(\sigma M_1 v_1^a +  \gamma M_2 v_2^a\right)/ M_{\rm crit} \, ,
\end{eqnarray}
and
\begin{equation}
h_-^a=  h_1^a-h_2^a \, , \qquad v_-^a = v_1^a - v_2^a\, .
\end{equation}
In terms of these new fields the quadratic Lagrangian 3-form takes the form ${\cal L}^{(2)}={\cal L}_+^{(2)} + {\cal L}_-^{(2)}$, with
\begin{eqnarray}
\!{\cal L}^{(2)}_+  &=& -M_{\rm crit}\left[ h_{+\, a} \bar{\cal D} v_+^a +
\frac{1}{2} \varepsilon_{abc} \bar e^a v_+^b v_+^c  \right. \nonumber \\
&& \left. \qquad\qquad  -\  \frac{1}{2} \Lambda \,  \varepsilon_{abc} \bar e^a h_+^b h_+^c \right]\, ,
\end{eqnarray}
where $\bar {\cal D}$ is the covariant exterior derivative with respect to the background, and
\begin{eqnarray}
{\cal L}^{(2)}_- &=& - \frac{\sigma\gamma M_1 M_2}{M_{\rm crit}} \left[ h_{-\, a} \bar{\cal D} v_-^a +
\frac{1}{2}\varepsilon_{abc} \bar e^a v_-^b v_-^c \right. \nonumber \\
&& \left. \qquad +\ \frac{1}{2} \left({\cal M}^2-\Lambda\right) \varepsilon_{abc} \bar e^a h_-^b h_-^c \right]\, ,
\end{eqnarray}
where
\begin{equation}\label{FPmass}
{\cal M}^2 = m^2 \left(\beta_1+ \gamma \beta_2\right)\frac{M_{\rm crit}}{\sigma M_1+M_2} \, .
\end{equation}

The form fields $v_\pm^a$ appearing in the above quadratic Lagrangian 3-forms  are auxiliary and may be eliminated by their field equations. Eliminating $v_+^a$ we find that ${\cal L}_+^{(2)}$ becomes the quadratic approximation to the Einstein-Hilbert Lagrangian density in the AdS background; this does not propagate any modes.
Eliminating $v_-^a$ we find that ${\cal L}_-^{(2)}$ is proportional to the  Fierz-Pauli Lagrangian density,  in the AdS background, for a spin-$2$ field with FP mass ${\cal M}$.
Notice that $\Lambda$ contributes to the mass term, which is zero when ${\cal M}^2=\Lambda$; this is the ``partially massless'' case where the linearised theory acquires an additional gauge invariance. This case is not relevant when  $\Lambda<0$ because there will be a spin-$2$ tachyon unless ${\cal M}^2>0$.
The parameters of the model are  further restricted by the requirement of positive kinetic energy, which amounts to bulk unitarity in the quantum theory;
recalling that $M_1,M_2>0$, we must have
\begin{equation}\label{unitarity}
\frac{\sigma\gamma }{M_{\rm crit}}>0 \, .
\end{equation}
Although the quadratic Lagrangian ${\cal L}^{(2)}$ is not diagonalisable for $M_{\rm crit}=0$, we can still take the limit $M_{\rm crit}\to 0$ in the field equations. The  massive modes become formally massless in this limit.
We shall not discuss this ``critical'' case here.

The fact that ZDG propagates just two physical  modes (which happen to be spin-$2$ modes) implies that the dimension, per space point, of the physical phase space of the linearized theory
is 4. This remains true in perturbation theory but does not exclude the appearance of additional degrees of freedom in other backgrounds. However, it is possible to determine the non-perturbative dimension of the physical phase space by Hamiltonian methods. As the action is Chern-Simons-like, being constructed as the integral  of products of forms without an explicit metric,  it is already first-order and a space/time split, e.g.~$e_\mu^a = \left(e_0^a , e_i^a\right)$ ($i=1,2$),  suffices to put it  into a form that is ``almost'' Hamiltonian, with the Hamiltonian being a sum of Lagrange multipliers times constraint functions. However, the field equations will  generically imply additional secondary constraints and these should be included too (there are no tertiary constraints if one starts from a first-order Chern-Simons-like action \cite{Routh:2013uc}). In the case of ZDG, there are two secondary constraints:
 \begin{eqnarray}\label{scon}
0 &=& \varepsilon^{ij} e_{1\, i} \cdot e_{2\, j} \, , \\
0 &=& \varepsilon^{ij} \left[ \beta_1 \left(\omega_1-\omega_2\right)_i \cdot e_{1\, j} +
  \beta_2 \left(\omega_1-\omega_2\right)_i \cdot e_{2\, j}\right]\, , \nonumber
 \end{eqnarray}
where the dot product  notation implies contraction of the 3-vectors with the Lorentz metric.  In the NMG limit (to be discussed below) these reduce to the two secondary constraints found for that model in \cite{Hohm:2012vh}.   Each  3-vector form field in the action adds $2\times 3=6$ (per space point)  to the phase-space dimension  (from its space components) and contributes three primary constraints (its time components  are the Lagrange multipliers). As we have four such  fields, the initial phase space has dimension 24 (per space point) and there are 12 primary constraints, to which we must add the two secondary constraints, making a total of 14 constraints; of these, six are first class (corresponding to the six EC gauge invariances) and 8 are second class. The constraints therefore reduce the phase space dimension by $2\times 6 +8 =20$, leaving a physical phase space of dimension $24-20=4$ (per space point), in agreement with  the linearised analysis. The counting here is exactly the same as that given for NMG in \cite{Hohm:2012vh} but the detailed verification of the fact that there are six first-class and eight second-class constraints (which we omit) is different.

Any 3D gravity model admitting an AdS  vacuum will also admit the asymptotically AdS black hole metric found
by Ba\~nados, Teitelboim and Zanelli (BTZ)  as solutions of 3D GR \cite{Banados:1992wn}.
Generically, the mass (and entropy) of BTZ black holes is positive whenever the central charge of the dual CFT is positive.
Therefore,
in the NMG model these BTZ black holes have negative mass whenever the bulk spin-$2$ modes have positive energy.
As we shall see, ZDG overcomes this problem.

The central charge of the boundary CFT for  GR follows directly from the results of  Brown and Henneaux on asymptotic symmetries in AdS$_3$  \cite{Brown:1986nw}; their result is  $c= 24 \pi \ell M_P$, where $\ell$ is the AdS$_3$ radius (so $\Lambda= -1/\ell^2$) and $M_P= 1/(16\pi G)$ where $G$ is the 3D Newton constant, which has dimensions of inverse mass in units for which the speed of light is unity.  A similar computation for ZDG, which we have verified using Hamiltonian methods, shows that
\begin{equation}\label{ccZDG}
c= 12\pi \ell\,  M_{\rm crit} \, \qquad ({\rm ZDG})\; .
\end{equation}
We note that for $M_2=0$ and $\sigma=1$ this reduces to the Brown-Henneaux result, using that in our normalization of (\ref{LI}) the Planck mass
is $M_P=M_1/2$.
It was to be expected that the central charge would be proportional to $M_{\rm crit}$ because it should vanish for the critical gravity case.

We are now in a position to determine whether there is a parameter range for ZDG for which perturbative unitarity in the bulk is compatible with positive central charge of the boundary
CFT.   When $\sigma=-1$ the bulk unitarity condition $-\gamma/M_{\rm crit} >0$ is incompatible with the condition $M_{\rm crit}>0$ unless $\gamma<0$, but then $M_{\rm crit}<0$ from its definition, so we require both $\sigma= 1$ and an AdS vacuum with $\gamma>0$ for compatibility of $c>0$ with bulk unitarity.  Absence of tachyons, ${\cal M}^2>0$, then requires that
\begin{equation}\label{notachyon}
\beta_1+\gamma \beta_2>0\, .
\end{equation}
Of course, this result applies only when there is an AdS vacuum, so we also need to check  that (\ref{quadgam}) allows a real positive solution for $\gamma$ such that $\Lambda<0$.

A simple explicit case satisfying all the above conditions can  found by setting
\begin{equation}
M_1=M_2\, , \quad \beta_I = 1\, , \quad \alpha_I = \frac{3}{2} + \zeta \, ,
\end{equation}
where $\zeta$ is a positive constant.  For this choice the quadratic equation (\ref{quadgam}) reduces to $\gamma^2=1$, and choosing $\gamma=1$ we get an AdS vacuum with
$ \left(\ell m\right)^{-2} = \zeta$.   Furthermore,  $\gamma \approx 1$ for any   `nearby' ZDG model, with slightly  different parameters, which are themselves constrained only by inequalities that have been satisfied but not saturated.  It follows that the above explicit model is one of an open set of models in the ZDG parameter space with  similar ``good'' properties; these properties are {\it not} the result of any fine-tuning of parameters that could be destabilized by perturbative quantum corrections. There could also be higher-derivative quantum corrections, of course,  but these can be dealt with, in perturbation theory, in the same way as in GR.

We conclude with a discussion of how the general ZDG model unifies previous models of massive 3D gravity. To see the relation to the 3D dRGT model we set $\sigma=1$ and write
\begin{equation}
e_2^a = \bar e_2^a + \lambda h_2^a\, , \qquad \omega_2^a = \bar \omega_2^a + \lambda v_2^a\, ,
\end{equation}
where $\bar e_2^a$ is a ``reference dreibein'' and $\bar\omega_2^a$ the corresponding (zero torsion) reference spin connection, and $\lambda$ is a constant.
If we now take $\lambda\to0$ keeping fixed $\lambda^2M_2 =M_1 \equiv 2 M_P$ then the ZDG Lagrangian 3-form reduces to
\begin{equation}
{\cal L} = 2 M_P {\cal L}_0(h_2,v_2) + 2 M_P {\cal L}_{\rm dRGT}(e_1,\omega_1)\, ,
\end{equation}
where the first term is the quadratic approximation  to the 3D EC  action (in the reference background)  for the fluctuations $(h_2^a,v_2^a)$; because of its linearised EC gauge invariances, this term propagates no modes. In the second term we have, after renaming $(e_1^a,\omega_1^a)$ as $(e^a,\omega^a)$,
\begin{eqnarray}
{\cal L}_{\rm dRGT}&=& -e_aR^a - \alpha_1\frac{m^2}{6} \varepsilon^{abc} e_a e_b e_c  \\
&& \ +\,  \frac{m^2}{2} \varepsilon^{abc}\left(\beta_1 e_a e_b \bar e_c + \beta_2 e_a \bar e_b \bar e_c\right) \, , \nonumber
\end{eqnarray}
which is the dreibein form of the 3D dRGT model.

To see the relation of ZDG to NMG we set $\sigma=-1$ and  write
\begin{equation}\label{NMGlim}
e_2^a = e_1^a + \frac{\lambda}{m^2} f^a\, , \quad \omega_2^a = \omega_1^a - \lambda h^a \, .
\end{equation}
We now consider the `flow'
 \begin{equation}\label{floW}
 \begin{split}
  M_1(\lambda) &= 2\Big(1+\frac{1}{\lambda}\Big)M_P\;, \quad M_2(\lambda) = \frac{2}{\lambda}M_P\;, \\
  \alpha_1(\lambda) &= -\frac{\lambda_0}{m^2}\lambda +\frac{1}{\lambda}\,, \qquad\; \alpha_2(\lambda) = 2\Big(1+\frac{1}{\lambda}\Big)\;, \\
  \beta_1(\lambda) &= 0 \;, \qquad\qquad\qquad \;\;\, \beta_2(\lambda) =  1\;,
 \end{split}
 \end{equation}
 and send $\lambda\rightarrow 0$ for fixed Planck mass $M_P$. This is the first-order formulation of the limit considered in \cite{Paulos:2012xe},  which exists only if the kinetic terms for $e_1^a$ and $e_2^a$ have opposite sign, i.e.~only if $\sigma=-1$.
Then  the $\lambda\to0$ limit of ${\cal L}_{\rm ZDG}$ exists too, with the final result that
\begin{eqnarray}
{\cal L} &=& 2M_P \left[ e_a R^a +  \frac{\lambda_0}{6} \varepsilon^{abc} e_a e_b e_c +  h_a T^a \right. \nonumber \\
&& \left. \qquad\quad   - \ \frac{1}{m^2} \left(f_a R^a + \frac{1}{2} \varepsilon^{abc} e_a f_b f_c\right)\right]\, .
\end{eqnarray}
This is the ``Chern-Simons-like'' action for NMG  \cite{Hohm:2012vh}. The second dreibein has now become an auxiliary field; by eliminating it we recover the higher-derivative action for  NMG.

As a consistency check we now verify that the central charge (\ref{ccZDG}) of ZDG reduces in the above limit to the
known central charge of NMG. From (\ref{fixLam}) and (\ref{quadgam}) we learn that
\begin{equation}
\Lambda(\lambda) = \Lambda_0+{\cal O}(\lambda)\, , \quad  \gamma(\lambda) = 1-\frac{\Lambda_0}{2m^2}\lambda +{\cal O}(\lambda^2)\, ,
\end{equation}
where $\Lambda_0$ is the cosmological constant in NMG as determined by the NMG field equations in terms of the cosmological parameter $\lambda_0$.
Now we insert (\ref{floW}) into (\ref{ccZDG}), use the NMG relation  $\lambda_0 = \Lambda_0^2/(4m^2) -\Lambda_0$ \cite{Bergshoeff:2009hq}, and then
write  $\Lambda_0=-1/\ell^2$  to deduce that
 \begin{equation}
  c(\lambda) \ = \ -24\pi\ell M_P\left(1-\frac{1}{2\ell^2m^2}\right)+{\cal O}(\lambda)\, .
 \end{equation}
The limit $\lambda\rightarrow 0$ indeed gives the NMG central charge.

We recall that NMG has a parity-violating extension to a  ``general massive gravity'' (GMG) model  in which the two spin-$2$ modes have unequal masses \cite{Bergshoeff:2009hq}. It  would be natural to suppose that ZDG is also a special case of a  more general parity-violating theory that propagates two massive spin-$2$ modes with unequal masses. By taking one mass to infinity we would then have a generalization of   ``topologically massive  gravity''  \cite{Deser:1981wh}  and hence of  ``chiral gravity''  \cite{Li:2008dq}. However, although it  is not difficult to find generalizations of ZDG  that have a limit to GMG,   those that we have considered  have additional degrees of freedom that only go away in the GMG  limit, so this remains an open problem.

In this paper we have presented a ``zwei-dreibein'' model of 3D massive gravity that both incorporates earlier parity-invariant models, such as NMG,  and extends them in such a way as to resolve the clash between bulk and boundary unitarity in the context of the AdS/CFT correspondence. Moreover, this is achieved without the need to fine tune parameters, so it is a result that is robust against quantum renormalization of parameters.  As both bulk and boundary unitarity are essential for  quantum consistency,
the model constructed here may be the first candidate for a semi-classical approximation to a consistent  quantum theory  of 3D massive gravity.
\vskip 0.1cm

\noindent\textbf{Acknowledgements}:  We thank Jan de Boer, Kyriakos Papadodimas,  Jan Rosseel, Alasdair Routh and Ivo Sachs for helpful discussions. EB, OH and WM thank the Galileo Galilei Institute for Theoretical Physics for hospitality during the initial stages of this work, and the organisers of the GGI program ``Higher Spins, Strings and Dualities'' for making this possible. OH is supported by the DFG Transregional Collaborative Research Centre and the DFG cluster of excellence ``Origin and Structure of the Universe''. SdH and WM are supported by the Dutch stichting voor Fundamenteel Onderzoek der Materie (FOM).

\centerline{\bf Erratum}
\smallskip

It was recently shown \cite{Banados:2013fda} that the ZDG model of massive gravity presented above 
has, for generic parameters,  an additional local degree of freedom in some backgrounds. The derivation, briefly presented above,  
of the secondary Hamiltonian constraints that exclude this possibility is not valid for generic parameters given the stated assumptions (invertibility of both dreibeine). To correct for this it is necessary to restrict the parameters such that 
$\beta_1\beta_2=0$; in the case that $\beta_1=0$, the required secondary constraints can be derived assuming only invertibility of $e_2$ (the second dreibein).  An explicit 
restricted ZDG model of this type that satisfies all the (bulk and boundary) unitarity conditions found above  is 
\begin{eqnarray}
\beta_1 &=& 0 \, , \quad \beta_2 = 1\,;  \qquad M_1=M_2 \, ; \nonumber \\
\alpha_1 &=& \frac12 + \zeta\,, \quad  \alpha_2 =  1  + \zeta \, \qquad \left(\zeta > 0\right)\,.
\end{eqnarray}
In this case the quadratic equation for $\gamma$, eqn.~(\ref{quadgam}) above, has the 
solution $\gamma=1$ and it yields an AdS vacuum with radius of curvature proportional to $\sqrt{\zeta}$.

\noindent\textbf{Acknowledgements}:  We thank M.~Ba\~nados, C.~Deffayet and M.~Pino for sending us an early version of their paper and for important discussions. %OH is supported by the DFG Transregional Collaborative Research Centre and the DFG cluster of excellence ``Origin and Structure of the Universe''. SdH and WM are supported by the Dutch stichting voor Fundamenteel Onderzoek der Materie (FOM).

\end{document}